\documentclass[12pt,nofootinbib]{revtex4-2}
\usepackage{amsfonts,amsmath,amssymb}
\usepackage{txfonts}
\usepackage{graphicx,color}
\usepackage{wrapfig,lipsum}
\usepackage{floatrow,float}
\usepackage{epstopdf}
\usepackage{hyperref}
\usepackage[normalem]{ulem}
\usepackage{enumitem}
\usepackage{mathrsfs,mathtools}

\begin{document}
	
\title{Grid Cells, Border Cells and Discrete Complex Analysis}
\author{Yuri Dabaghian}
\affiliation{Department of Neurology, The University of Texas McGovern Medical School, 6431 Fannin St, Houston, TX 77030\\
	$^{*}$e-mail: Yuri.A.Dabaghian@uth.tmc.edu}
\vspace{17 mm}
\date{\today}
	
\begin{abstract}
	We propose a mechanism enabling the appearance of border cells---neurons firing at the boundaries of the
	navigated enclosures. The approach is based on the recent discovery of discrete complex analysis on a
	triangular	lattice, which allows constructing discrete epitomes of complex-analytic functions and making
	use of their inherent ability to attain maximal values at the boundaries of generic lattice domains. As it
	turns out, certain elements of the discrete-complex framework readily appear in the oscillatory models of
	grid cells. We demonstrate that these models can extend further, producing cells that increase their activity
	towards the frontiers of the navigated environments. We also construct a network model of neurons with 
	border-bound firing that conforms with the oscillatory models. 
	%These results demonstrate that the maximum	principle may drive the physiological computations producing
	%border cell spiking patterns.
\end{abstract}

%{\bf Keywords:} grid cells, border cells, percolation, discrete complex analysis, learning and memory, hippocampo-cortical network

\maketitle
	
\section{Introduction and motivation}
\label{sec:int}

\begin{figure}[h]
	\centering
	\includegraphics[scale=0.817]{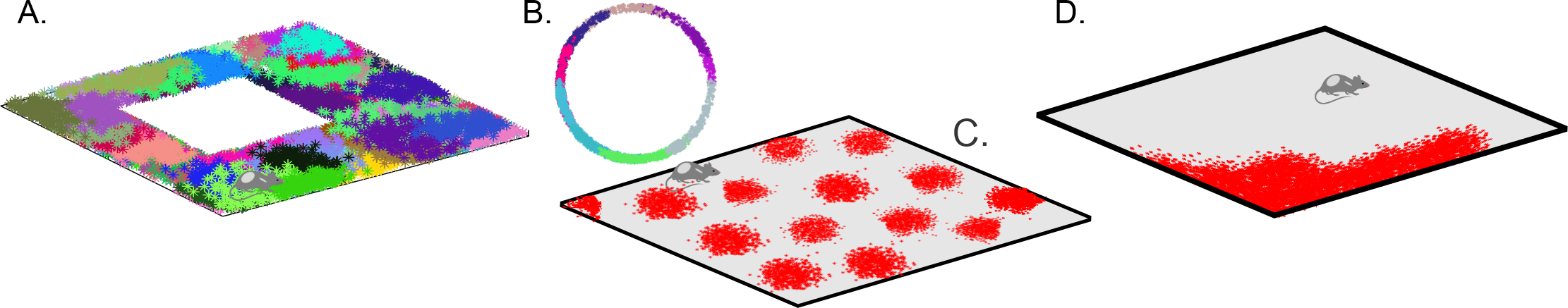}
	\caption{{\footnotesize
			\textbf{Spatial cells}. (\textbf{A}). Spikes produced by place cells (dots of different colors)
			form distinct spatial clusters in the navigated environment, which highlight the preferred spiking
			domains---place fields.
			\textbf{B}. Head direction cells fire when the animals' head is oriented at a particular angle with
			respect to cardinal directions, thus producing spike clusters in the circular space of planar 
			directions.
			\textbf{C}. Spiking domains of the grid cells form a triangular lattice that tiles the ambient space.
			\textbf{D}. Boundary cells produce spikes along the border of the navigated enclosure.
	}}
	\label{fig:gc}
\end{figure}

Spiking activity of spatially tuned neurons is believed to enable spatial cognition \cite{MoserRev,Grieves,
Derdikman}. For example, rodent's \textit{place cells}\footnote{Throughout the text, terminological definitions
and semantic highlights are given in \textit{italics}.} that fire in specific locations produce a qualitative 
map of the explored environment \cite{Goth,Alvrh,eLife,Wu,Rueck}; \textit{head direction} cells that fire each
at its preferred orientation of the animals' head contribute directional information \cite{Taube2,Aff,Valerio};
the \textit{grid cells} that fire near vertexes of a planar triangular lattice are believed to provide a metric
scale \cite{Haft,MosM} and the \textit{border cells} highlight the boundaries of the navigated enclosures 
\cite{LeverB,BarryB,SolstadB} (Fig.~\ref{fig:gc}A).

A number of theoretical models aim to explain the machinery producing these spiking profiles, by exploiting 
suitable mathematical phenomena, e.g., attractor network dynamics \cite{Tsodyks,RollsAtt,ColginAtt,Bassett,
	Giocomo}, specific network architectures \cite{ColginAtt,Bush,ChFr,SolstM}, the hexagonal symmetry of
closely packed planar discs \cite{Fuhs}, constructive interference of symmetrically propagating waves
\cite{OscG1,OscG2,OscG3,OscG4} and so forth.
In contrast, the ability of border cells to identify the frontiers of the explored environments was heretofore
explained heuristically, as a  certain ``responsiveness'' these neurons to the walls of the navigated arenas, achieved, 
conceivably, by integrating proprioceptive and sensory  inputs \cite{GDet1,GDet2,Raudies,HartleyB,BurgessB}.
However, since border cells are anatomically removed from sensory pathways, it is possible that their spiking 
may be produced through autonomous network mechanisms, rather than induced by external driving. 
From a computational perspective, such mechanisms may also hinge on a mathematical phenomenon that highlights
the perimeters of spatial regions, a well known example of which is the \textit{maximum principle}---the ability
of certain functions, e.g., harmonic and complex-analytic functions, to attain maximal values at the boundaries
of their domains \cite{Marsden}. 
%%%%%%%%%%%%%%%%%%%%%%%%%%%%%%%%%%%%%%%%%%%%%%%%%%%%%%%%%%%%%%%%%%%

%%%%%%%%%%%%%%%%%%%%%%%%%%%%%%%%%%%%%%%%%%%%%%%%%%%%%%%%%%%%%%%%%%%
The following study is motivated by a recent series of publications \cite{ND1,SPN1,SPN2,IDynn}, which show that
two-dimensional ($2D$) triangular lattices allow constructing a discrete counterpart of the Complex Analysis
and defining real-valued, discrete epitomes of complex-analytic functions that obey the maximum principle. As
it turns out, these structures allow modeling border cell activity, as discussed below. 

The paper is organized as follows. Several key ideas of Discrete Complex Analysis (DCA) are outlined in 
Section~\ref{sec:app}, following the exposition given in \cite{ND1,SPN1,SPN2,IDynn}. Section~\ref{sec:grid}
discusses certain connections between elements of DCA and oscillatory interference models of grid cells
\cite{OscG1,OscG2,OscG3,OscG4}, and offers a generalized framework for expanding these models to include border
cell spiking patterns. In Section~\ref{sec:res}, elements of DCA are implemented in a schematic network model
that produces border cell firing responses through endogenous activity, without using external parameters, such
as animal's speed or location. The results are briefly discussed in Section.~\ref{sec:disc}.

\section{Approach}
\label{sec:app}

\textbf{1. Discrete complex analysis}. Standard theory of complex variables is a calculus over complex numbers
$z=x+iy$ and their conjugates, $\bar{z}=x-iy$, where $x$ and $y$ are the Cartesian coordinates in a Euclidean 
plane and $i$ is the imaginary unit, $i^2 =-1$ \cite{Marsden}. 
A generic complex function depends on both $z$ and $\bar{z}$; however, the main objects of the theory are the
\textit{analytic} (also called \textit{holomorphic}) functions that depend only on $z$, $f=f(z)$, and their 
\textit{anti-analytic} (\textit{anti-holomorphic}) counterparts, that depend only on $\bar{z}$, $f=f(\bar{z})$. 
The defining property of these functions is that their derivatives over the ``missing" variable vanish,
\begin{equation}
		\begin{aligned}
		\frac{\partial f}{\partial\bar{z}}&=\left(\frac{\partial}{\partial x}+i\frac{\partial}{\partial y}
		\right)f=0, \,\,\,\, \mbox{for analytic functions}, & \\ 
		\frac{\partial f}{\partial\bar{z}}&=\left(\frac{\partial}{\partial x}-i\frac{\partial}{\partial y}
		\right)f=0, \,\,\, \mbox{for anti-analytic functions}. & 
	\end{aligned}
		\nonumber
\end{equation} 

The Cauchy operator and its conjugate used above,
\begin{equation*}
	\partial \equiv\frac{\partial}{\partial x}+i\frac{\partial}{\partial y}, \,\,\,\,\,\,
	\bar{\partial}\equiv\frac{\partial}{\partial x}-i\frac{\partial}{\partial y},
\end{equation*} 
play key roles not only in complex analysis but also in geometry and applications. One of their properties is
that they factorize the $2D$ Laplace operator, or the \textit{Laplacian},
\begin{equation}
	\Delta \equiv\frac{\partial^2}{\partial x^2}+\frac{\partial^2}{\partial y^2}=
	\left(\frac{\partial}{\partial x}+i\frac{\partial}{\partial y}\right)
	\left(\frac{\partial}{\partial x}-i\frac{\partial}{\partial y}\right)\equiv\partial\bar{\partial}.
	\label{ddbar}
\end{equation}
The factorization (\ref{ddbar}) is unique and necessarily involves complex numbers---think of the decomposition
$x^2+y^2=(x+iy)(x-iy)$ that is commonly used to motivate the transition from real to complex variables. 
Correspondingly, the phenomenon (\ref{ddbar}) takes place only on spaces that admit complex 
structure---orientable $2D$ surfaces. Furthermore, the factorization (\ref{ddbar}) can serve as a vantage 
point for defining the Cauchy operator and its conjugate: if a Laplacian admits the decomposition (\ref{ddbar})
in suitable coordinates, then the resulting curvilinear first-order operators $\bar{\partial}$ and $\partial$
will be the Cauchy operators of a complex-analytic structure on the corresponding manifolds. 

\begin{wrapfigure}{c}{0.6\textwidth}
	\centering
	\includegraphics[scale=0.92]{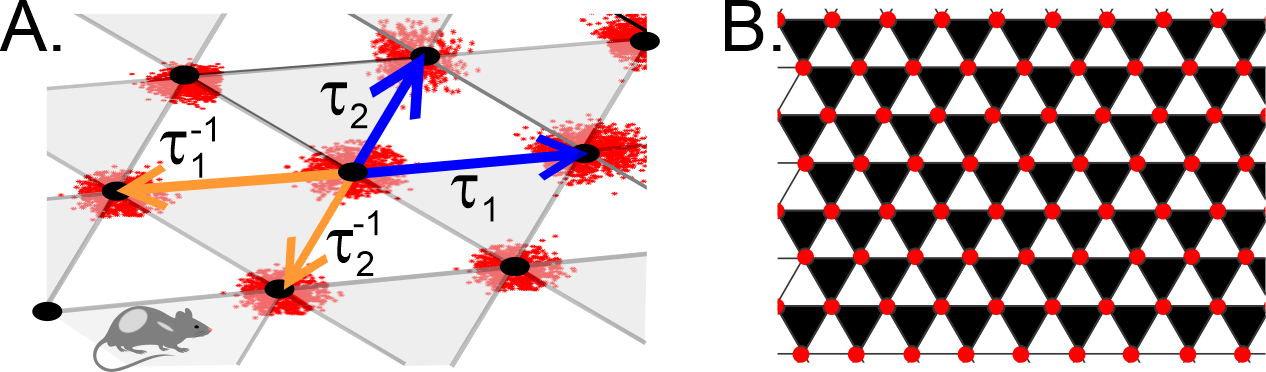}
	\caption{\footnotesize{
			\textbf{Lattice}. \textbf{A}. The operators $\tau_1$ and $\tau_2$ (blue arrows), shift the argument
			of the lattice function forward along the basis directions, from $(m,n)$ to $(m+1,n)$ and $(m,n+1)$
			respectively. The inverse operators, $\tau_1^{-1}$ and $\tau_2^{-1}$ (orange arrows), shift the
			argument backwards, correspondingly to $(m-1,n)$ and $(m,n-1)$.
			\textbf{B}. The backwards shifts $\tau_1^{-1}$ and $\tau_2^{-1}$ support the ``black" triangles and
			forward shifts $\tau_1$ and $\tau_2$ span the complimentary set of ``white" triangles. If a function
			satisfies the discrete-analyticity condition (\ref{qbarf}), then its values over the black triangles
			vanish.}
	}
	\label{fig:lttc}
\end{wrapfigure}

A remarkable observation made in \cite{ND1,SPN1,SPN2,IDynn} is that the \textit{discrete} Laplace operator
on a $2D$ triangular lattice also is factorizable. Indeed, a generic discrete Laplacian on a graph or a 
lattice acts on the vertex-valued functions $f(v)$ as
\begin{equation}
	\Delta f(v)=\sum_{v'}f(v')-\rho_{v}f(v),
	\label{DiscLap}
\end{equation}
where the summation goes over all vertexes $v'$ linked to $v$, and $\rho_v$ is the valency of $v$ \cite{Godsil,DLap,Sarnak}.
On a triangular lattice with vertexes marked by two integer indexes $m$ and $n$, the Laplacian (\ref{DiscLap}) 
becomes
\begin{equation}
	\Delta f =f(m+1,n+1)+f(m+1,n)+\ldots+f(m-1,n-1)-6f(m,n).
	\label{DiscLapL}
\end{equation}
To obtain the required decomposition, let us define the operators $\tau_1$ and $\tau_2$ that shift the arguments
of the vertexes functions,
\begin{subequations}
	\label{taur}
	\begin{align}
		\tau_1f(m,n)&=f(m+1,n),\tag{4a}\\
		\tau_2f(m,n)&=f(m,n+1).\tag{4b}
	\end{align}
\end{subequations}
as shown on Fig.~\ref{fig:lttc}A. In terms of $\tau_1$ and $\tau_2$, the sum (\ref{DiscLapL}) becomes
\begin{equation}
	\Delta_L=\tau_1+\tau_2+\tau_1^{-1}+\tau_2^{-1}+\tau_2\tau_1^{-1}+\tau_1\tau_2^{-1}-6,
	\label{lap}
\end{equation}
and factorizes into the product of two first-order operators 
\begin{subequations}
\label{Qr}
	\begin{align}
		Q &=1+\tau_1+\tau_2,\label{Q}\tag{6a}\\
		\bar{Q} &=1+\tau_1^{-1}+\tau_2^{-1},\label{Qbar}\tag{6b}
	\end{align}
\end{subequations}
with an extra constant term,
\begin{equation}
	\Delta_L =Q\bar{Q}-9.
	\label{QQ}
\end{equation}
As shown in \cite{ND1,SPN1,SPN2,IDynn}, this decomposition induces a DCA, in which the operator $\bar{Q}$
plays the role of the complex-conjugate derivative $\bar{\partial}$. One can thus define the discrete-analytic
lattice functions, $f(m,n)$, as the ones that satisfy the relationship
\begin{subequations}
	\label{qq}
	\begin{align}
		\begin{split}
			\bar{Q}\,f(m,n) =f(m,n)+f(m-1,n)+f(m,n-1) =0.\label{qbarf}
		\end{split}
		\tag{$8a$}
		\intertext{The $Q$-operator then acts as the discrete-analytic derivative,}
		\begin{split}
			Q\,f(m,n) =f(m,n)+f(m+1,n)+f(m,n+1). \label{qf}
		\end{split}
		\tag{$8b$}
	\end{align}
\end{subequations}
Geometrically, equations (\ref{qq}) can be illustrated by partitioning the lattice $V$ with ``black" and 
``white" triangles, in which each white triangle, $\vartriangle$, shares sides with three black triangles,
$\blacktriangledown$, and vice versa (Fig.\ref{fig:lttc}B). According to (\ref{qbarf}), the discrete analytic
functions vanish over all the black triangles, which may be viewed as the lattice analogue of 
``$z$-but-not-$\bar{z}$" dependence of the conventional complex-analytic functions.
%%%%%%%%%%%%%%%%%%%%%%%%%%%%%%%%%%%%%%%%%%%%%%%%%%%%%%%%%%%%%%%%%%%

%\begin{figure}[h]
%\end{figure}

%%%%%%%%%%%%%%%%%%%%%%%%%%%%%%%%%%%%%%%%%%%%%%%%%%%%%%%%%%%%%%%%%%%
\textbf{2. Properties of the discrete-analytic functions} largely parallel the familiar properties of their
continuous counterparts, including the maximal principle that is used below to model the border cell spiking
activity.  However, there are also a few differences, the most striking of which is that the discrete-analytic
functions are \textit{real-valued}: indeed, the equation (\ref{qbarf}) does not involve imaginary numbers and
possesses real-valued solutions \cite{ND1,SPN1,SPN2,IDynn}. Thus, the discrete complex analysis is a 
real-valued combinatorial framework that may be implemented through neuronal computations\footnote{DCA can also
be constructed over the complex numbers: the corresponding theory then yields the standard Complex Analysis in
the limit when the lattice side vanishes \cite{ND1,SPN1,SPN2,IDynn}.}.

\begin{figure}[h]
	\centering
	\includegraphics[scale=0.84]{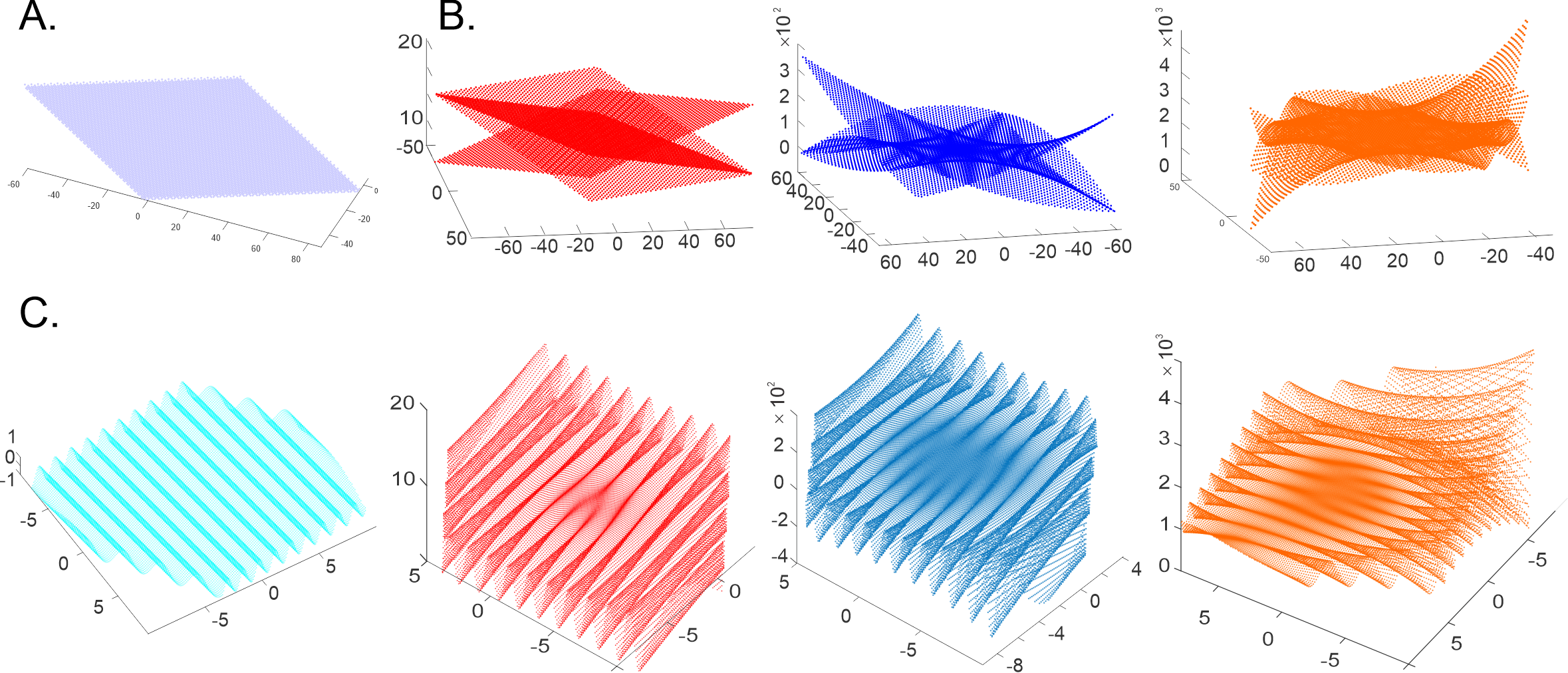}
	\caption{\footnotesize{
		\textbf{Discrete-analytic polynomials}. \textbf{A}. $0^{\textrm{th}}$ order polynomials are the 
		holomorphic constants that assume a small set of discrete values $h_1=\{-0.5,1\}$ and $h_2=\{\pm
		\sqrt{3}/2,0\}$.
		\textbf{B}. The discrete-holomorphic polynomials $P_1(m,n)$,  $P_2(m,n)$ and $P_3(m,n)$ grow outwards
		(linearly, parabolically and cubically) as the	lattice indexes increase.
		\textbf{C}. The spatially-refined discrete polynomials produce undulatory shapes scaffolded by their
		discrete counterparts: shown are the undulating holomorphic wave $h_1(x,y)$ and the polynomials $P_1
		(x_1,x_2)$,	$P_2(x_1,x_2)$ and $P_3(x_1,x_2)$ that grow towards the boundary of the enclosed Euclidean
		domain.
	}}
	\label{fig:Pn}
\end{figure}

Another peculiarity is that DCA redefines the notion of a constant. Indeed, the constants $c$ of the standard 
calculi are nullified by the derivatives, $\partial c=\bar{\partial}c=0$. However, a quantity
that assumes constant values on all vertexes, $f(m,n)=c$, is not nullified, but tripled by discrete derivative
operators, $Qc=\bar{Q}c=3c$. Hence, discrete-analytic constants $h$ must be derived from the equations
\begin{equation}
	\bar{Q}h=Qh=0.
	\label{qqh}
\end{equation}
Somewhat surprisingly, the basic solutions of (\ref{qqh}) have the form
\begin{equation}
	h(\delta)=\cos\frac{2\pi}{3}(n+2m+\delta),
	\label{const1}
\end{equation} 
where $\delta$ is a phase parameter (Fig.~\ref{fig:Pn}A). Formula (\ref{const1}) can be viewed as a discrete
analogue of the complex phase $e^{i\delta}$; the ``prime" constants $1$ and $i$ then correspond to
\begin{subequations}
	\label{const}
	\begin{align}
		h_1=\cos\frac{2\pi}{3}(n+2m),\label{h1}\tag{11a}\\
		h_2=\sin\frac{2\pi}{3}(n+2m).\label{h2}\tag{11b}
	\end{align}
\end{subequations}
Note that, in contrast with their familiar counterparts, the ``constants" (\ref{const1}) and (\ref{const})
alternate from vertex to vertex, assuming a few discrete values, $h_1=\{-0.5,1\}$ and $h_2=\{\pm\sqrt{3}/2,0\}$.

The third distinct property concerns Taylor-expansions: in contrast with the continuous case, a generic
discrete-holomorphic function $f(m,n)$ over a finite lattice domain can be represented exactly by finite
series, i.e., one can write
\begin{equation*}
	f(m,n)= U(m,n)h_1+W(m,n)h_2,
\end{equation*}
where $U(m,n)$ and $W(m,n)$ are polynomials.  The order of such polynomials generally grows with the size of
the lattice domain, which allows keeping the above expansion exact.

Explicit examples of the first, second and third-order discrete-analytic polynomials are
\begin{subequations}
	\label{ph123}
	\begin{align}
		P_1 & =-\frac{\sqrt{3}}{2}(m+n)h_1+\frac{1}{2}(n-m)h_2,\label{polyn1}\vspace{6pt}\tag{12$a$}\\[12pt]
		P_2 &=(m-n)(3(n+m)-1)h_1-\sqrt{3}((m+n)^2+2mn-3(n+m))h_2.\label{polyn2}\tag{12$b$}\\[12pt]
		P_3 &=(m-n)((m+2n)(2m+n)-2(3(m+n)-1))h_1\label{polyn3}\tag{12$c$}\\
		&+\sqrt{3}(6(m+n)-2mn-4(m+n)^2+3mn(m+n))h_2,
		\nonumber
	\end{align}
\end{subequations}
illustrated on Fig.~\ref{fig:Pn}B, C and D. It can be verified by direct substitution\footnote{The operators $Q$
and $\bar{Q}$ generally do not distribute according to the Leibniz rile, e.g., $$Q(f(m,n)h_2) \neq(Qf(m,n))h_1
+f(m,n)(Qh_2).$$} that the operator $\bar{Q}$ nullifies each polynomial, whereas $Q$ lowers their order, $QP_1
=h_1$, $QP_2\propto P_1$ and $QP_3\propto P_2$, just as $\bar{\partial}$ would nullify polynomials of $z$, and
$\partial$ would lower their order, $\partial p_r(z)\propto p_{r-1}(z)$. 
In general, there are $2(r+1)$ basic discrete-analytic polynomials of order $r$, which corresponds to $2(r+1)$
basic complex $r^{\textrm{th}}$-order complex polynomials \cite{ND1,SPN1,SPN2,IDynn}.
%%%%%%%%%%%%%%%%%%%%%%%%%%%%%%%%%%%%%%%%%%%%%%%%%%%%%%%%%%%%%%%%%%%

%%%%%%%%%%%%%%%%%%%%%%%%%%%%%%%%%%%%%%%%%%%%%%%%%%%%%%%%%%%%%%%%%%%
\textbf{3. Spatial fine-graining.} Discrete functions defined over the lattice vertexes give rise to 
finer-grained spatial structures. Given two basis vectors  
\begin{equation}
	\vec{e}_1^{\,g}=a_g(1,0),\qquad \vec{e}_2^{\,g}=a_g\left(1/2,\sqrt{3}/2\right),
	\label{basis}
\end{equation} 
in the Euclidean plane, consider a lattice generated by integer translations,
\begin{equation}
	V_g=\{v_{m,n}^g=m \vec{e}_1^{\,g}+n \vec{e}_2^{\,g}, \,\, m,n\in\mathbb{Z}\}.
	\label{vmn}
\end{equation}
Such embedding allows extending the discrete argument of a vertex function, $f(m,n)$, to a function of Euclidean
 coordinates, $f(x_1,x_2)$, by replacing the integer arguments $(m,n)$ with
pairs of reals $(x_1,x_2)$. For example, the discrete-holomorphic constant (\ref{h1}) yields a continuous 
``holomorphic wave'' with wavelength $\propto a_g$, 
\begin{equation}
	\cos\frac{2\pi}{3}(2m+n)\to\cos\frac{2\pi}{3a_g}(2x_1+x_2),
	\label{grain}
\end{equation}
propagating in the direction $\vec{e}_1$ (Fig.~\ref{fig:Pn}C). Conversely, using
\begin{equation*}
	x_1=a_g m+\delta_1, \,\,\,\,\,\, x_2=a_g n+\delta_2
\end{equation*}
in the real-valued functions with sufficiently low spatial frequency (less than $2\pi/a_g$) restores the 
dependence upon the lattice indexes and produces a continuous phase $\delta$ that contains fractional 
remainders,
\begin{equation}
	\cos\frac{2\pi}{3a_g}(2x_1+x_2)\to\cos\frac{2\pi}{3}(n+2m+\delta).
	\label{holwave}
\end{equation}
The latter form allows acting with the operators $Q$ and $\bar{Q}$ on the regular coordinate functions and
placing the results into the context of DCA.

\section{Oscillatory Grid cell models}
\label{sec:grid}

\begin{figure}[h]
		\centering
		\includegraphics[scale=0.8]{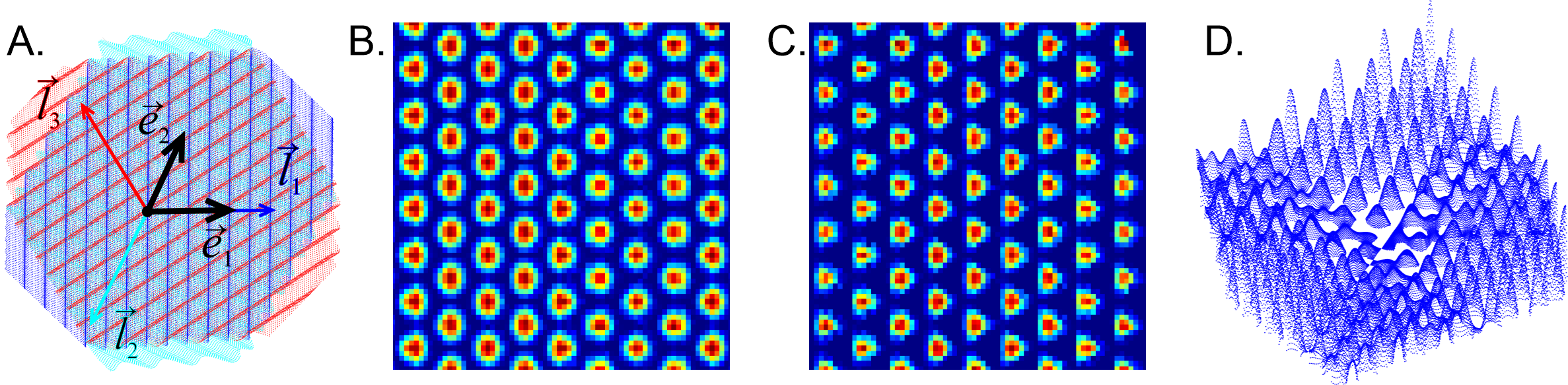}
		\caption{\footnotesize{
			\textbf{Oscillatory interference model}. \textbf{A}. Superposition of three discrete-holomorphic
			waves propagating in three symmetric directions specified by the three wave vectors $\vec{l}_1$,
			$\vec{l}_2$ and $\vec{l}_3$. The basis lattice directions $\vec{e}_1$ and $\vec{e}_2$ are shown
			in black. Constructive interference occurs at the vertexes of a triangular lattice, highlighted
			by the amplitude (\ref{Ag}).
			\textbf{B}. The grid cell firing amplitude, $A_g$, formula (\ref{3cos}), reproduces the familiar
			grid cell layout. \textbf{C}. The complementary ``conjugate" amplitude, $\tilde{A}_g=\sin\varphi_1
			\sin\varphi_2\sin\varphi_3$.
			\textbf{D}. The second order discrete-holomorphic \textit{grid-polynomial} (\ref{pg2}), defined
			over the grid fields, compare with the third panel on Figs.~\ref{fig:Pn}C.
			}
		}
		\label{fig:PnW}
\end{figure}

Surprisingly, discrete-analytic structures are manifested in the existing models of grid cell activity, e.g.,
in the oscillatory interference models that derive the observed grid field patterns from the dynamics of the
membrane potential,
\begin{equation}
	\mu_g(t)=\prod_{k=1}^{3}\left(\cos(\omega t)+\cos\left(\omega t+\beta\int_0^{t}\langle\vec{l}^{\,g}_{k}
	\cdot\vec{v}\rangle dt\right)\right)_{\theta}.
	\label{memb}
\end{equation}
Here $t$ is time, $\beta$ is a scale parameter, $\vec{l}^{\,g}_1$, $\vec{l}^{\,g}_2$ and $\vec{l}^{\,g}_3$
are the three symmetric wave vectors, $\vec{v}(t)$ is the velocity, and $\omega\approx 8$ Hz is the frequency
of synchronized extracellular field's oscillations. The index ``$\theta$" refers to the firing threshold 
\cite{OscG1,OscG2,OscG3,OscG4}. Due to the symmetry, the waves interfere constructively at the vertexes of a
triangular lattice with basis vectors $\vec{e}_1^{\,g}=\vec{l}^{\,g}_1$ and $\vec{e}_2^{\,g}=-\vec{l}^{\,g}_2$,
centered at the firing fields\footnote{The model \cite{Fuhs} uses sum of three waves for capturing analogous
	interference effect.} (Fig.~\ref{fig:gc}C, \ref{fig:PnW}A).

To link $\mu_g(t)$ to DCA, let us rewrite the time integrals in (\ref{memb}) as integrals along the trajectory,
\begin{equation*}
	\mu_g(t)=\prod_{k=1}^{3}\left(\cos\omega t+\cos\left(\omega t+\frac{8\pi}{3a_g}\langle\vec{l}_k^{\,g}\cdot
	\int_{\gamma}d\vec{r}\rangle\right)\right)_{\theta}=A_g(\vec{r})\prod_{k=1}^{3}\cos\left(\omega t+\varphi
	_k^g(\vec{r})\right),
\end{equation*}
where $\vec{r}$ is the position vector, $\dot{\vec{r}}=\vec{v}$, $\varphi_k^g$ are the oscillatory phases and
$8\pi/3a_g=\beta$. The time-independent factor defines the spatial amplitude of the membrane potential,
\begin{equation}
	A_g(\vec{r})=\prod_{k=1}^{3}\cos\frac{4\pi}{3a_g}\left(\vec{l}^{\,g}_k\cdot\vec{r}\right)_{\theta},
	\label{Ag}
\end{equation}
and produces the familiar spatial pattern of grid fields, brought about by the constructive interference of the
contributing waves (Fig.~\ref{fig:PnW}B). Next, given the rat's position in the lattice basis, $\vec{r} = m
\vec{e}_1^{\,g} + n \vec{e}_2^{\,g}+\delta\vec{r}$ and using $\vec{l}_3^{\,g}=\vec{e}_2^{\,g}-\vec{e}_1^{\,g}$,
yields
\begin{eqnarray}
	A_g(\vec{r})=\cos\frac{2\pi}{3}(2m+n +\delta_1)\cos\frac{2\pi}{3}(2n+m +\delta_2)\cos\frac{2\pi}{3}
	(n-m +\delta_2-\delta_1),
	\label{3cos}
\end{eqnarray}
where $\delta_k=2\delta\vec{r}\cdot\vec{e}_k^{\,g}$ are the remainder phases. Curiously, each multiplier in
(\ref{3cos}) is a discrete-holomorphic constant: the second coincides with (\ref{const1}), the first can be 
obtained from (\ref{const1}) by re-indexing, $m\leftrightarrow n$, and the last is produced by an index shift,
$n\to n-3m$. Even more surprisingly, the full product (\ref{3cos}), adjusted by a constant reference value 
$1/4$, is also nullified by the discrete Cauchy operators,
\begin{equation*}
	Q\left(A_g-1/4\right)=\bar{Q}\left(A_g-1/4\right)=0,
\end{equation*}
which means that the amplitude of grid cells' firing (\ref{Ag}) is, in fact, a basic DCA object---a fine-grained
discrete-holomorphic constant that functionally highlights the lattice of firing fields. The neurons that 
respond to grid cell outputs can hence be viewed as functions on that lattice, which includes 
discrete-holomorphic functions used for modeling border cells. Furthermore, the necessary elements of the DCA
can be constructed independently within the oscillatory model, as discussed below.
%%%%%%%%%%%%%%%%%%%%%%%%%%%%%%%%%%%%%%%%%%%%%%%%%%%%%%%%%%%%%%%%%%%

%%%%%%%%%%%%%%%%%%%%%%%%%%%%%%%%%%%%%%%%%%%%%%%%%%%%%%%%%%%%%%%%%%%
\section{Border cells}
\label{sec:res}
\textbf{Oscillatory model} of the grid cells can be generalized to simulate border cells' activity by replacing
the constant membrane potential (\ref{memb}) with suitable discrete-holomorphic functions obeying the maximum
principle. The resulting firing rate will then grow towards the boundary of the navigated environment 
$\mathcal{E}$ and produce the characteristic border cell firing patterns. 

A simple implementation of this idea can be achieved using the discrete-analytic polynomials (\ref{ph123}),
by replacing the combinations
\begin{equation*}
		\theta_1=2m+n,\hspace{10pt}
		\theta_2=m+2n,\hspace{10pt}
		\theta_3=n-m, %\hspace{10pt}
\end{equation*}
with the phases appearing in (\ref{Ag}),
\begin{equation*}
	\theta_i\to\varphi_i\equiv\frac{2\pi}{3a_g}\langle\vec{l}_i^{\,g}\cdot\vec{r}\rangle,
\end{equation*}
that represent dendritic inputs into the postsynaptic cell \cite{Almeida}. The resulting fine-grained
discrete-holomorphic polynomials are then
\begin{subequations}
	\label{pr123}
	\begin{align}
		P_1^h &=-\frac{\sqrt{3}}{3}\varphi_{12}h_1-\frac{1}{2}\varphi_3 h_2,
		\label{pln1}\tag{20$a$}\\
		P_2^h &=\varphi_3(2\varphi_{12}-1)h_1-\sqrt{3}\left(\frac{4}{6}\varphi_{12}^2-\frac{1}{2}\varphi_3^2
		-2\varphi_{12}\right)h_2,
		\label{pln2}\tag{20$b$}\\
		P_3^h &=\varphi_3(\varphi_1\varphi_2-2(2\varphi_{12}-1))h_1+\sqrt{3}\left(4\varphi_{12}
		-2\varphi_{12}^2+\frac{2}{9}\varphi_{12}^3-\frac{1}{2}\varphi_3^2(\varphi_{12}-1)\right)h_2,
		\label{pln3}\tag{20$c$}
		\nonumber
	\end{align}
\end{subequations}
where $\varphi_{12}$ is a short notation for $(\varphi_1+\varphi_2)/2$ and the waves $h_1$, $h_2$ in 
(\ref{ph123}) can be steered along any of the symmetric directions, $\vec{l}_1$, $\vec{l}_2$ or $\vec{l}_3$.

Physiologically, it is possible\footnote{Currently, the synaptic organization of the border cell network is
debated.} that border cell activity is gated by inputs from the grid cells \cite{Ktz,Flor,Gisig,Hayman,GioAnc,
Rowland}. This mechanism can be modeled by replacing the ``undulating" holomorphic constants $h_1$ and $h_2$ in
(\ref{ph123}) with the grid cell firing amplitudes, $A_g$ and the complementary combination of holomorphic sine
waves $\tilde{A}_g=\sin\varphi_1\sin\varphi_2\sin\varphi_3$ (Fig.~\ref{fig:PnW}C), which yields \textit{grid 
	polynomials}, e.g.,
\begin{subequations}
	\label{pg123}
	\begin{align}
		P_1^g &=-\frac{\sqrt{3}}{3}\varphi_{12}A_g-\frac{1}{2}\varphi_3 \tilde{A}_g,
		\label{pg1}\tag{21$a$}	\\
		P_2^g &=\varphi_3(2\varphi_{12}-1)A_g-\sqrt{3}\left(\frac{4}{6}\varphi_{12}^2-\frac{1}{2}\varphi_3^2
		-2\varphi_{12}\right)\tilde{A}_g,
		\label{pg2}\tag{21$b$}\\
%		P_3^g &=\varphi_3(\varphi_1\varphi_2-2(2\varphi_{12}-1))A_g+\sqrt{3}\left(4\varphi_{12}
%		-2\varphi_{12}^2+\frac{2}{9}\varphi_{12}^3-\frac{1}{2}\varphi_3^2(\varphi_{12}-1)\right)\tilde{A}_g,
%		\label{pg3}\tag{21$c$}
		&\textrm{etc}., \nonumber
	\end{align}
\end{subequations}
defined explicitly over the grid field lattice (Fig.~\ref{fig:PnW}D). By direct verification, both sets of 
polynomials (\ref{pr123}) and (\ref{pg123}) are discrete-analytic functions that obey the maximum principle 
and can hence serve as building blocks for producing generic membrane potentials accumulating towards the
boundaries of the navigated enclosures.

\begin{figure}[h]
		\centering
		\includegraphics[scale=0.845]{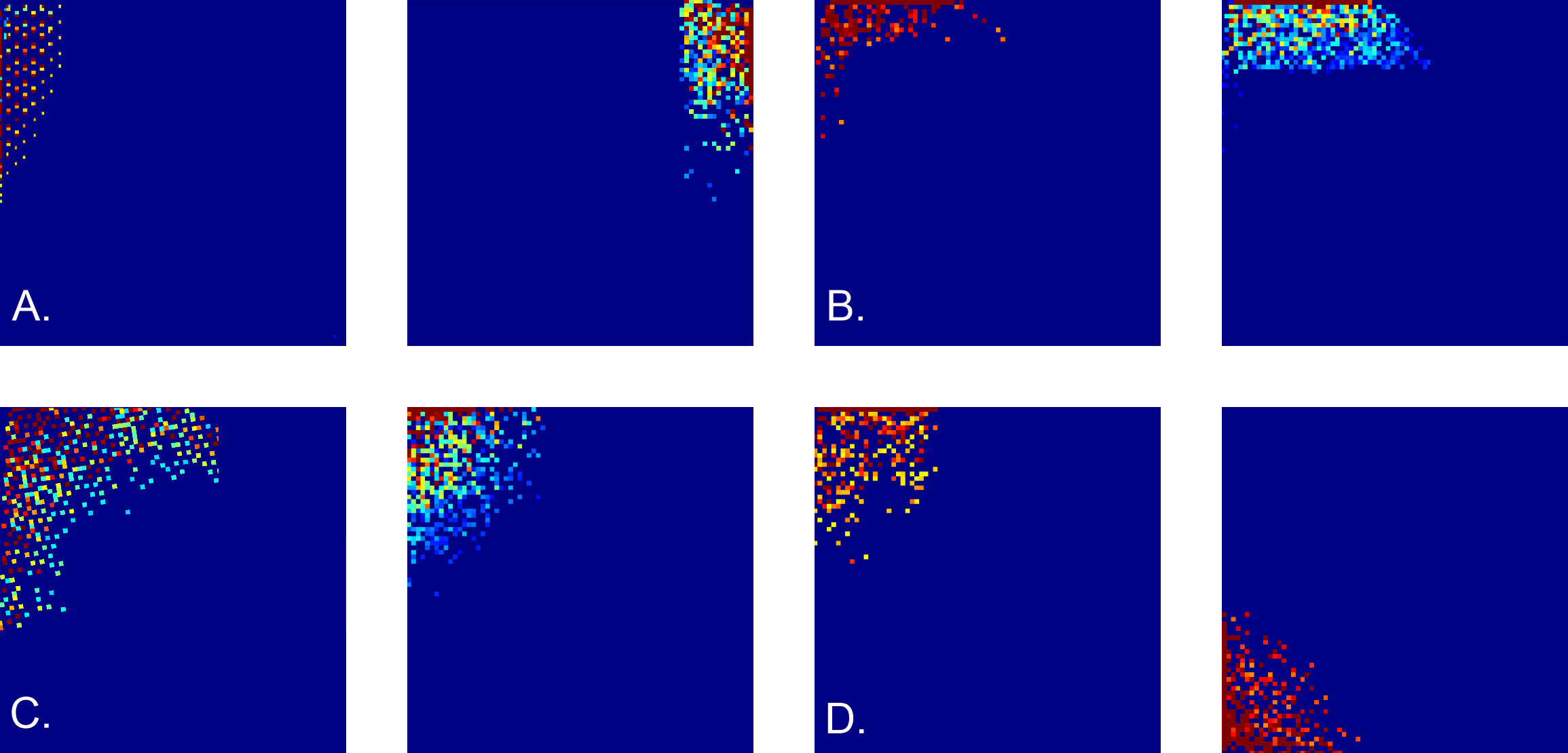}
		\caption{\footnotesize{
				\textbf{Border cell firing patterns}. Examples of simulated border cell firing fields in a
				square $6\times 6$ m environment, obtained as combinations of first, second, and third order
				grid polynomials. \textbf{A}. Firing fields obtained using ``undulating" discrete-holomorphic
				polynomials (\ref{ph123}). 
				\textbf{B}. Examples of the firing fields obtained using combinations of grid-holomorphic 
				polynomials (\ref{pg123}).
				\textbf{C}. Firing fields of noise-perturbed membrane potentials, for $\varepsilon=0.2$ (left 
				panel) and  $\varepsilon=0.3$ (right panel).
				\textbf{D}. Firing fields obtained using the schematic network model.
			}
		}
		\label{fig:BC}
\end{figure}

As mentioned above, the individual $\varphi$-terms in (\ref{pr123}) and (\ref{pg123}) may be physiologically 
interpreted as the inputs received through linear or nonlinear synapses. Since the second- and the third-order
nonlinear synapses are discussed in the literature \cite{Rajan1,Rajan2,Liu,Latimer,Mahes,Brivio,Bicknell,Biane,
	Todo,Wang,Rossbroich}, we used combinations of $5$-$10$ polynomials of the orders $r_i=1,2,3$,
\begin{equation}
	\mu_b=\left[\alpha_1 P^{\ast}_{r_1}+\alpha_2 P^{*}_{r_2}+\ldots+\alpha_q P^{*}_{r_q}\right]_{\theta}.
	\label{gb}
\end{equation}
Here the $P^{*}_{r}$ represent either harmonic (\ref{pr123}) or the grid polynomials (\ref{pg123}), the 
coefficients $\alpha_i$ define the magnitude of each addend, and the $\theta$ subscript indicates the threshold.
In the simulations, the values $\alpha_i$ were selected randomly, while the threshold grew according to the 
size of the environment and the order of the contributing polynomials, $\theta\propto(L/a_g)^{r_i}$. The
resulting firing maps are illustrated on Fig.~\ref{fig:BC}A,B. Expectedly, since all contributing polynomials in
(\ref{gb}) grow towards the boundaries of the available lattice domain, all simulated border cells fire along 
the frontiers of the navigated enclosure.
%%%%%%%%%%%%%%%%%%%%%%%%%%%%%%%%%%%%%%%%%%%%%%%%%%%%%%%%%%%%%%%%%%%

%%%%%%%%%%%%%%%%%%%%%%%%%%%%%%%%%%%%%%%%%%%%%%%%%%%%%%%%%%%%%%%%%%%
Importantly, these outcomes are robust with respect to stochastic variations: disturbing the phases $\varphi_i$
of the holomorphic polynomials with a noise term, $\varepsilon\xi$, where $\xi$ is a random variable uniformly
distributed over $[0,2\pi]$ and $\varepsilon$ controls its amplitude, does not qualitatively alter the resulting
spatial patterns for $\varepsilon\leq0.5$ or more (Fig.~\ref{fig:BC}C).

\textbf{Schematic network model}. Defining the membrane potentials as functions of speed and coordinates used,
e.g., in (\ref{memb}) helps linking the geometry of the observed environment to the underlying neuronal 
computations. However, modeling the brain's own representation of the ambient environment requires using
intrinsic representation of spatial information, a key role in which is played by hippocampal place cells,
$c_i$, and the postsubicular\footnote{Head direction cells are also found in few other brain regions 
	\cite{Taube2}.} head direction cells, $h_i$ \cite{Grieves,Taube2}. The computational units enabling 
this representation are the functionally interconnected cell groups 
\begin{subequations}
	\label{cqs}
	\begin{align}
		\sigma_i&=[c_{i_0},c_{i_1},\ldots, c_{i_n}], \label{sigma}\tag{23$\sigma$}\\
		\eta_j&=[h_{j_1},h_{j_2},\ldots,h_{j_n}],\label{eta}\tag{23$\eta$}
	\end{align}
\end{subequations}
%%%%%%%%%%%%%%%%%%%%%%%%%%%%%%%%%%%%%%%%%%%%%%%%%%%%%%%%%%%%%%%%%%%
which highlight, respectively, basic locations $\upsilon_{\sigma_i}$ and angular domains $\upsilon_{\eta_j}$ 
\cite{Harris2,Peyrache1,Peyrache2,BrandonTh,Maurer}. A number of studies have demonstrated that the assemblies
(\ref{cqs}) encode the animal's ongoing position, the shape of trajectory and even its planned and recalled 
navigational routes \cite{Brown,Frank,Guger,Karlsson1,Johnson,Dragoi,Pfeiffer1}. By the same principle, place
cell assemblies that fire over the grid fields $\upsilon_{g_i}$, can provide their hippocampal representation:
a combination $\hat{\sigma_i}$ of $\sigma$-assemblies whose constituent cells exhibit coactivity with a grid
cell $g$ and each other defines a vertex of grid cell activity,
\begin{equation}
	\hat{\varv}_{i}^g=[\hat{\sigma}_{i},g].
	\label{v}
\end{equation} 
In the following, the superscript ``$g$" will be suppressed in describing single grid cell activity and used 
only to distinguish contributions from different grid cells.

The hexagonal order on the vertexes (\ref{v}) is established by concomitant activity of select groups of head
direction assemblies, $\hat{\eta}_{1},\hat{\eta}_{2},\ldots,\hat{\eta}_{6}$, that activate on the runs between
pairs of neighboring grid fields, e.g., $\upsilon_i$ and $\upsilon_j$, thus defining the \textit{spiking edges}
between $\hat{\varv}_i$ and $\hat{\varv}_j$,
\begin{equation}
	\epsilon_{ij}^{k}=\{\hat{\sigma}_i,\hat{\sigma}_j|\hat{\eta}_{k},g\}.
	\label{e}
\end{equation}
Together, the vertexes (\ref{v}) and the edges (\ref{e}) can be viewed as elements of a \textit{spike-lattice}
$\mathcal{V}_g$, by which the grid field lattice is embedded in the cognitive map \cite{Percol}. Using
$\mathcal{V}_g$ allows constructing a self-contained phenomenological network model of border cells that does
not involve ``tagging" the neuronal activity by externally observed characteristics, such as the rat's speed or
Euclidean coordinates. 

Suppose that a cell $b$ with membrane potential $\mu_b$ receives input from a group of persistently firing
head direction assemblies $\hat{\eta}_k$, over a period when grid cell $g$ becomes active, then shuts down,
and then restarts its activity again\footnote{For a physiological discussion, see 
	\cite{Hslm1,Hslm2,Egorov,Percol}.}.
If these consecutive activations are induced over adjacent vertexes $\hat{\varv}_i$ and $\hat{\varv}_j$, then
the corresponding change of the membrane potential can be interpreted as the change of the spike-lattice 
function $\mu_b(\hat{\varv})$ along the edge $\epsilon_{ij}$ between them,
\begin{equation}
	[\hat{\sigma}_i,\hat{\sigma}_j,\hat{\eta}_k, g]\rightsquigarrow\mu_b(\hat{\varv}_i)=\mu_b(\hat{\varv}_j).
	\label{fv}
\end{equation}
On the other hand, the transformation (\ref{fv}) can be described as the action of a \textit{spike-lattice 
	shift operator} $\hat{\tauup}$ on $\mu_b$, 
\begin{equation*}
	\hat{\tauup}\mu_b(\hat{\varv}_i)=\mu_b(\hat{\varv}_j).
\end{equation*} 
In particular, changes induced by the head direction assemblies $\hat{\eta}_1$ and $\hat{\eta}_2$ (ordered as
on Fig.~\ref{fig:lttc}A) can be identified with the shift operators acting ``forward'' along the basic lattice
directions,
\begin{subequations}
	\label{tt1}
	\begin{align}
		\begin{split}
			\hat{\tauup}_1 \mu_b(\hat{\varv})&= \mu_b(\hat{\varv}_{+}')\,\,\, \textrm{and} \,\,\, \hat{\tauup}_2
			\mu_b(\hat{\varv})= \mu_b(\hat{\varv}_{+}''),
			\label{tau}
		\end{split}
		\tag{27a}
		\intertext{while the ``opposite'' assemblies $\hat{\eta}_4$ and $\hat{\eta}_5$ induce backward
			transformation,}
		\begin{split}
			\hat{\tauup}_1^{-1} \mu_b(\hat{\varv})&= \mu_b(\hat{\varv}_{-}')\,\,\, \textrm{and} \,\,\, 
			\hat{\tauup}_2^{-1} \mu_b(\hat{\varv})= \mu_b(\hat{\varv}_{-}'').
			\label{tau1}
		\end{split}
		\tag{27b}
	\end{align}
\end{subequations}

The appearance of spiking analogues of the shift operators $\tau_1$ and $\tau_2$ associated with grid cells
opens a possibility of implementing the key DCA structures neuronally. However, a principal challenge in this
approach is that the series of inputs received along a particular trajectory may not concur with the lattice 
structure of the underlying grid fields. Indeed, consider the membrane potential at the initial spiking vertex
$\hat{\varv}_0$,
\begin{equation*}
	\mu_b(\hat{\varv}_0)= U(\hat{\varv}_0)A_g(\hat{\varv}_0)+W(\hat{\varv}_0)\tilde{A}_g(\hat{\varv}_0),
\end{equation*} 
from where the animal continues to move along a trajectory $\gamma$, producing a series of postsynaptic
changes described by a sequence of $\hat{\tauup}$-shifts,
\begin{equation}
	\mu_b(\hat{\varv}_{f})=\hat{\tauup}_{i_1}\hat{\tauup}_{i_2}\ldots\hat{\tauup}_{i_k}\cdot(U(\hat{\varv}_0)
	A_g(\hat{\varv}_0)+W(\hat{\varv}_0)\tilde{A}_g(\hat{\varv}_0)).
	\label{vv}
\end{equation} 
If the net membrane potential (\ref{vv}) does not depend on the order in which the individual inputs arrive,
the $\hat{\tauup}$-operators commute\footnote{As do their $\tau$-counterparts \cite{ND1,SPN1,SPN2,IDynn}.}.
Thus, the value accrued at the final vertex $\hat{\varv}_{f}$ is
\begin{equation}
	\mu_b(\hat{\varv}_{f})\equiv\mu_b(\hat{\varv}_{\tilde{m},\tilde{n}})=\hat{\tauup}_1^{\tilde{m}}\hat{\tauup}
	_2^{\tilde{n}} \cdot(U(\hat{\varv}_0)A_g(\hat{\varv}_0)+W(\hat{\varv}_0)\tilde{A}_g(\hat{\varv}_0)),
	\label{vf}
\end{equation}
where the integers $\tilde{m}$ and $\tilde{n}$ mark how many times $\hat{\tauup}_1^{\pm}$ and $\hat{\tauup}_2^
{\pm}$ were triggered along the way. 
%%%%%%%%%%%%%%%%%%%%%%%%%%%%%%%%%%%%%%%%%%%%%%%%%%%%%%%%%%%%%%%%%%%

\begin{figure}[h]
		\centering
		\includegraphics[scale=0.775]{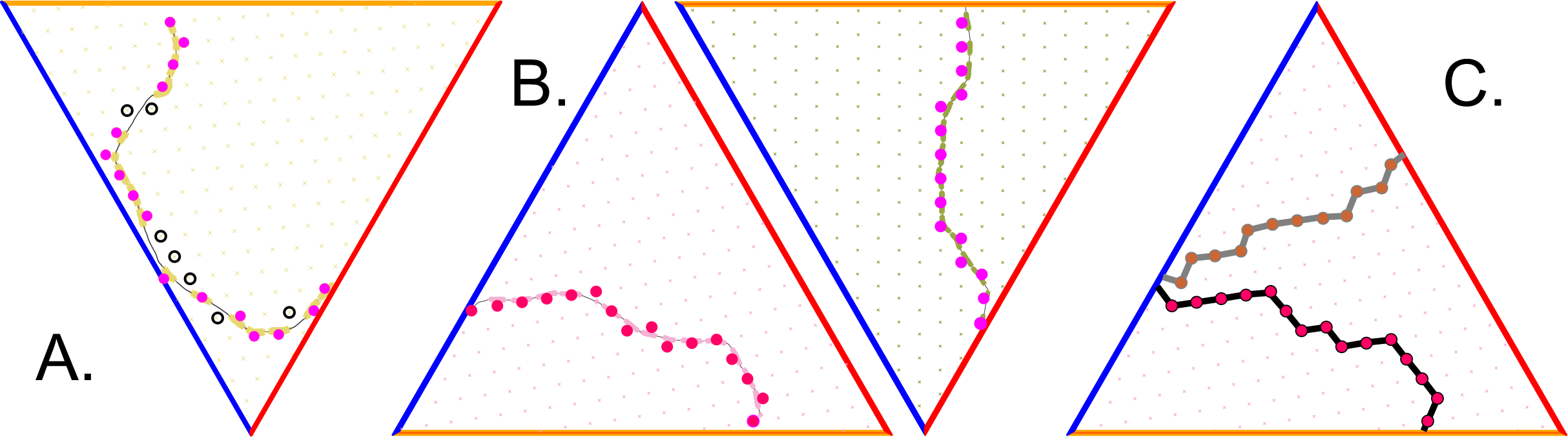}
		\caption{\footnotesize{
				\textbf{Grid cell percolation and border cell firing patterns}. 
				\textbf{A}. Generic path is non-percolative: the vertexes that correspond to the ``percolated"
				firing fields---the	ones over which the grid cell has produced at least one spike---are	marked
				by pink, while vertexes	corresponding to the  fields that did not respond are marked by black.
				\textbf{B}. Two examples of percolating trajectories, along which the spiking occurred at each
				vertex, without omissions.  
				\textbf{C}. Two examples of lattice paths induced by two percolating trajectories.
		}
		}
		\label{fig:pr}
\end{figure}

%%%%%%%%%%%%%%%%%%%%%%%%%%%%%%%%%%%%%%%%%%%%%%%%%%%%%%%%%%%%%%%%%%%
Note however, that a generic trajectory $\gamma$ may not pass through the fields of a given cell $g$ in 
complete sequence: some fields are visited, others are occasionally missed (Fig.~\ref{fig:pr}A). As a result,
the ``empirical" $(\tilde{m},\tilde{n})$-indexing appearing in (\ref{vf}) may not conform with the original 
$(m,n)$-indexing of the full grid field set, which moots the possibility of interpreting the argument of 
$\mu_b$ in terms of the underlying lattice (\ref{vmn}). However, it can be shown that, within physiological
parameter range, there typically exists a special class of ``percolating" paths---those that run through the
firing fields of a given grid cell in contiguous sequence, without omissions (see \cite{Percol} and 
Fig.~\ref{fig:pr}A). Such paths induce series of conjoint spiking edges,
\begin{equation}
	\mathfrak{G}(\gamma)\equiv\{\epsilon_{ij},\epsilon_{jk},\ldots,\epsilon_{pq}\},
	\label{lpath}
\end{equation}
that serve as lattice representations of the animals' moves (Fig.~\ref{fig:pr}C, \cite{Percol}). The increments
of the postsynaptic membrane potential (\ref{vf}) acquired along the link series (\ref{lpath}) are, by design,
compatible with the lattice indexing and hence allow constructing consistent lattice functions over an extended
lattice domains \cite{Percol}. The subsequent development of the model will therefore be based on percolating
paths only.

Constructing a membrane potential (\ref{vf}) by applying spiking $\hat{\tauup}$-operators along the percolated
paths requires knowing how these operators act on discrete-holomorphic constants and polynomials, which can be
established as follows. First, the response of the spike-lattice counterparts of holomorphic constants $h_1,
h_2$, and of their grid analogues, $A_g(\hat{\varv})$ and $\tilde{A}_g(\hat{\varv})$, to $\hat{\tauup}$-shifts
(\ref{tt1}), can be implemented according to how the corresponding original, index-dependent expressions
(\ref{const}) and (\ref{3cos}) respond to the $\tau$-operators, e.g.,
\begin{subequations} 
	\label{tA}
	\begin{align}
		\hat{\tauup}_1^{\pm1}A_g(\hat{\varv})&=-\frac{1}{2}A_g(\hat{\varv})
		\pm\frac{\sqrt{3}}{2}\tilde{A}_g(\hat{\varv}),\label{tA1}\tag{31.1}\\
		\hat{\tauup}_2^{\pm1}\tilde{A}_g(\hat{\varv})&=-\frac{1}{2}\tilde{A}_g(\hat{\varv})
		\mp\frac{\sqrt{3}}{2}A_g(\hat{\varv}).\label{tA2}\tag{31.2}
	\end{align}
\end{subequations}
%up to a constant. 
One can then use the expressions (\ref{tA}) along with (\ref{tt1}) as the rules defining how the $\hat{\tauup}$s
act on the spike-lattice $\mathcal{V}_g$, and thus deduce how the ``spiking" Cauchy operator $\bar{\mathcal{Q}}$
acts on generic membrane potentials,
\begin{eqnarray}
	\bar{\mathcal{Q}}\mu_b(\hat{\varv})=
	\left(U(\hat{\varv})-\frac{1}{2}U(\hat{\varv}_{-}')-\frac{1}{2}U(\hat{\varv}_{-}'')
	+\frac{\sqrt{3}}{2}(W(\hat{\varv}_{-}')-W(\hat{\varv}_{-}''))\right)
		A_g(\hat{\varv})\nonumber\\
	+\left(W(\hat{\varv})-\frac{1}{2}W(\hat{\varv}_{-}')-\frac{1}{2}W(\hat{\varv}_{-}'')
	+\frac{\sqrt{3}}{2}(U(\hat{\varv}_{-}')-U(\hat{\varv}_{-}''))\right)
	\tilde{A}_g(\hat{\varv}).
	\label{Qbarf}
\end{eqnarray}
To satisfy the  discrete analyticity condition, $\bar{\mathcal{Q}}\mu_b(\hat{\varv})=0$, the coefficients in 
front of the holomorphic constants in (\ref{Qbarf}) must vanish at each spike-vertex $\hat{\varv}$. The simplest
solution to  this requirement is provided by the functions that acquire constant increments over the vertex 
shifts,
\begin{subequations}
	\label{P1}
	\begin{align}
			U(\hat{\varv}_{\pm}')&=U(\hat{\varv})\pm C_1, \,\,\,
			U(\hat{\varv}_{\pm}'')=U(\hat{\varv})\pm C_2, \label{U1}\tag{33u}\\
			W(\hat{\varv}_{\pm}')&=W(\hat{\varv})\pm D_1, \,\,\, W(\hat{\varv}_{\pm}'')=W(\hat{\varv})\pm D_2. 
			\label{W1}\tag{33w}
		\end{align}
\end{subequations}
By direct verification, the equation $\bar{\mathcal{Q}}\mu_b(\hat{\varv})=0$ is satisfied identically if
\begin{equation}
	C_2	=-C_1=C, \,\,\,\,\,\, D_1=D_2=\sqrt{3}C,
	\label{PC1}
\end{equation}
where $C$ represents vertex-independent additive synaptic input. Thus, if the specific synaptic responses to
each of the $\tauup$s are defined by (\ref{PC1}), then the net accumulated postsynaptic membrane potential is
\begin{equation}
	\mu_b=C(m-n)\tilde{A}_g-C\sqrt{3}(m+n)A_g,
	\label{U1spk}
\end{equation}
which matches the linear discrete-holomorphic polynomial (\ref{pg1}) and clarifies how such potential may emerge
through synaptic integration. For the nonlinear membrane potentials described by higher-order polynomials, the
shifting rules can be obtained by analogy with (\ref{P1}), by requiring that the shifted values are described
by lower-order polynomials, e.g., by linear increments to the shifted second-order polynomials,
\begin{subequations}
	\label{P2}
	\begin{align}
		\Delta U_2(\hat{\varv})&=U_1(\hat{\varv})+C_1',\,\,\,\,\,\, \Delta U_2(\hat{\varv})=U_1(\hat{\varv})+C_2',
		\label{u2}\tag{36u}\\
		\Delta W_2(\hat{\varv})&=W_1(\hat{\varv})+D_1',\,\,\,\,\, \Delta W_2(\hat{\varv})=U_1(\hat{\varv})+D_2',
		\label{w2}\tag{36w}
	\end{align}
\end{subequations}
and so forth. The results then produce second and third order expressions of the type (\ref{pln2}) and 
(\ref{pg123}\textit{b}), which combine according to (\ref{gb}) and yield build border cell firing patterns as
illustrated on Fig~\ref{fig:BC}D.

\section{Discussion}
\label{sec:disc}

A number of computational models aim to explain the origins of the triangular spatial pattern of the grid cells' 
spiking activity and the contribution that these cells make into enabling spatial cognition \cite{Giocomo}.
It is believed that the regular grid firing patterns allow establishing global metric scale in the navigated
environment \cite{MosM} and may produce a spatial location code \cite{Welinder,Sreenivasan,Burak}. 
The model discussed above shows that the neuronal mechanisms producing hexagonal layout of the firing fields
may enable yet another mathematical phenomenon---a discrete complex structure. Although the whole structure
is implemented via real-valued computations, it captures all the key attributes of the conventional theory of
complex variables. In particular, the discrete-analytic functions defined in DCA framework obey the maximum
principle---a property that may be used to model neurons with firing responses tuned to the boundaries of the
navigated environments.

Surprisingly, basic elements of DCA are manifested in several existing models of grid cells. As discussed above,
the interfering waves of the oscillatory models, which may be viewed either as representation of physiological 
rhythms, or as formal spatiotemporal components of the membrane potential's  decomposition, can be described
as spatially fine-grained discrete holomorphic constants. Their net interference pattern, that defines the grid
cells' firing amplitude, also amounts to a discrete-holomorphic ``grid'' constant, highlighting a triangular
lattice. Incorporating higher-order polynomials (and hence generic discrete-holomorphic functions) into the 
model allows simulating border cell spiking, which emphasizes affinity of the two firing mechanisms. The latter
may explain why these cells are anatomically intermingled---in electrophysiological recordings, both cell types
are often detected on the same tetrode. According to the model, physiologically similar neurons may be wired to
perform synaptic integrations of different orders, and may, conceivably, swap their firing types in response to
synaptic or structural plasticity changes. 

The DCA approach can be also be used to produce self-contained network models that do not require 
phenomenological inputs, i.e., do not reference speed, coordinates, grid field positions, \textit{ad hoc} lattice
indexes $(m,n)$ or other externally observed tags of neuronal activity. On the contrary, it becomes possible to
render certain abstract DCA structures via autonomous network computations. For example, the Cauchy operators
and the lattice (\ref{vmn}) underlying the grid field layouts are induced using the ``spiking" analogues of the
$\tau$-operators (\ref{taur}),
\begin{equation}
	\mathcal{V}_g=\{\hat{\varv}_{m,n}=m\hat{\tauup}_1 +n \hat{\tauup}_2, \,\, m,n\in\mathbb{Z}\},
	\label{vg}
\end{equation}
with vertex indexes derived from counting synaptic inputs of the grid, head direction and place cells along the
percolated paths. In this context, the standard procedure of constructing grid fields $\upsilon_{m,n}^g$ 
(Fig.~\ref{fig:gc}C), by attributing $(x_1,x_2)$ coordinates to spikes according to the rat's ongoing location,
can be viewed as a  mapping from the vertexes of the spike lattice (\ref{vg}) into regions in the navigated 
environment,
\begin{equation*}
	f_g: \hat{\varv}_{m,n}^g\to \upsilon_{m,n}^g\in \mathcal{E},
	\label{f}
\end{equation*} 
centered at the vertexes of the grid field lattice $V_g$ \cite{Aff,SchemaS}. Zero holonomy property of the
discrete Cauchy operators discussed in \cite{ND1,SPN1} (see also \cite{HolReplay}) ensures that the $(m,n)$ 
values attained at a particular vertex do not depend on the percolating paths leading to a vertex, but only on
the vertex itself, which ensures consistency of the construction. The discrete-complex structure can thus be
viewed as an intrinsic network property, that may be implemented using different synaptic architectures, e.g.,
the continuous attractor models. An implication of this property is that the grid cells should be expected to
produce planar, rather than voluminous firing fields, in order to implement the Cauchy decomposition (\ref{QQ})
attainable only on $2D$ hexagonal lattices---a prediction that agrees with both experimental
\cite{Slope,Grieves3D,Aniso,Soman,Ginosar} and theoretical \cite{Stella,Mathis,Horiuchi,Gong} studies.

As a concluding comment, the DCA framework currently does not offer a direct geometric interpretation of the
discrete-holomorphic mappings \cite{ND1,SPN1,SPN2}. An independently developed notion of discrete conformal
transformations, based on rearrangements of regular circle packings in planar domains \cite{Koebe,Thurston,
	Bucking1,Rodin} may therefore offer a complementary venue for establishing correspondences between network
activity and discrete-complexity. Several recent experimental \cite{SavelliB,SavelliC,ZhangB,Krupic1,Krupic2,
	Wernle,Bell} and theoretical \cite{Spalla,Santos,Monsalve,Urda,Sharpee} studies suggest that conformal 
transformations of the navigated spaces may induce compensatory discrete-conformal transformations of the grid
field maps, similar to how the hippocampal place cells tend to preserve coactivity patterns in morphing 
environments \cite{eLife,Rueck,Goth}. If the latter is verified experimentally, it can be argued that the grid
cell inputs constrain the hippocampal topological map \cite{eLife,Rueck}, to a net conformal map of the 
navigated space.

\textbf{Acknowledgments}.

Work was supported by NSF grant 1901338 and NIH grant R01NS110806.

\end{document}